\documentclass[showpacs,amssymb,eqsecnum,twocolumn,aps]{revtex4}
\usepackage{amsmath}
\usepackage{graphicx} 
\usepackage{epsf}
\usepackage{graphics}
\usepackage[latin1]{inputenc}
\newcommand{\be}{\begin{equation}}
\newcommand{\ee}{\end{equation}}
\newcommand{\bea}{\begin{eqnarray}}
\newcommand{\eea}{\end{eqnarray}}
\newcommand{\undertilde}[1]{\underset{\widetilde{}}{#1}}

\newcommand{\unf}{\undertilde{F}}
\newcommand{\ung}{\undertilde{G}}
\newcommand{\una}{\undertilde{A}}
\newcommand{\unm}{\undertilde{M}}
\newcommand{\unn}{\undertilde{N}}
\newcommand{\unl}{\undertilde{\Lambda}}
\newcommand{\uns}{\undertilde{\Sigma}}

\begin{document}

\title{Non-Quadratic Gauge Fixing and Global Gauge Invariance in the Effective Action}

\author{F. T. Brandt$^{a}$  
 and D. G. C. McKeon$^{b}$} 
\affiliation{$^a$ Instituto de F\'{\i}sica,
Universidade de S\~ao Paulo,
S\~ao Paulo, SP 05315-970, Brazil}
\affiliation{$^b$ Department of Applied Mathematics, University of
Western Ontario, London, ON  N6A 5B7, Canada}

\begin{abstract}
The possibility of having a gauge fixing term in the effective
Lagrangian that is not a  quadratic expression  has been explored in
spin-two theories so as to have a propagator  that is both traceless
and transverse.  We first show how this same approach can be  used in 
spontaneously broken gauge theories as an alternate to the 't Hooft 
gauge fixing which avoids terms quadratic in the scalar fields. This
``non-quadratic'' gauge fixing  in the effective action results in
there being two complex Fermionic and one real Bosonic ghost fields.
A global gauge invariance involving a Fermionic gauge parameter,
analogous to the usual BRST invariance, is present in this effective action.  
\end{abstract}


\maketitle

\section{Introduction}
The path integral quantization of gauge theories involves having to choose a
gauge fixing condition  that results in a ``gauge fixing''  term
supplementing the classical Lagrangian. A further ``ghost action''
involving a complex Fermionic ghost field is also required in order to
cancel the contribution of non-physical degrees of freedom in
radiative process \cite{Feynman:1963ax,DeWitt:1967yk,Faddeev:1967fc,Mandelstam:1968hz}.
One of the most remarkable outcomes of this conventional gauge fixing
is that the full effective action possesses a global gauge invariance with a
Fermionic gauge parameter known as ``BRST invariance'' 
\cite{Tyutin:1975qk,Iofa:1976je,Becchi:1976nq}. 

In Ref. \cite{Brandt:2007td} it is pointed out how this procedure 
cannot be applied in the case of a spin-two gauge theory so as to have
a spin-two propagator that is both  traceless and transverse
(TT). This difficult can be circumvented by modifying the
``Faddeev-Popov'' procedure so that a ``gauge fixing'' 
term appears in the effective Lagrangian that is not quadratic in some
gauge fixing condition. 
In the next section we review this
procedure and provide a second illustration of how it works in a $U(1)$ model
in which there is spontaneous symmetry breaking. 

The gauge fixing term will still have the advantage of the 't Hooft gauge
\cite{tHooft:1971rn,Fujikawa:1972fe}, which eliminates the terms bilinear in the gauge
and scalar fields (which would appear if a Feynman gauge were used
\cite{Lee:1971kj}), and has the additional advantage of eliminating 
gauge dependent mass terms for the scalars which arise in the `t Hooft
gauge. The nature of the ghost fields which arise with this gauge
fixing is then discussed; it is shown that two Fermionic and one
Bosonic ghost arise. 

Having illustrated how non-quadratic gauge fixing terms can be
incorporated into the effective action, we now show how this effective
action possess a global gauge invariance which is a generalization of
the BRST invariance of the usual effective action. Both the Fermionic
and Bosonic ghosts participate in this transformation, with the
unusual feature that the transformation of the Fermionic ghost is non-local.


\section{Non-Quadratic Gauge Fixing}
We consider a vector gauge field interacting with a complex scalar 
field with classical Lagrangian
\bea\label{eq1}
{\cal L}_{Cl}&=&-\left(\partial_{\mu}+
ieV_\mu\right)\phi^\star\left(\partial_{\mu}-ieV_\mu\right)\phi
\nonumber \\ 
&\!\!\!\!\!\!-&\!\!\!\!\!\!
\frac{1}{4}\left(\partial_\mu V_\nu-\partial_\nu V_\mu\right)^2
-m^2\phi^\star \phi-\lambda (\phi^\star \phi)^2.
\eea
If $m^2<0$, spontaneous symmetry breakdown occurs, and if 
${f}$ is the vacuum expectation value of $\phi$,  so that
\be\label{eq2}
\sqrt{2} \phi = f + h
\ee
where $h$ is the quantum fluctuation about ${\cal C}$, then the model of
\eqref{eq1} develops a contribution that is bilinear in $V_\mu$ and
$h$. If the usual Feynman gauge fixing Lagrangian
$-\frac{1}{2\alpha}\left(\partial\cdot V\right)^2$ is used, these
terms result in a mixed  $\langle h V_\mu\rangle$ propagator which
complicates the computation of radiative effects. In order to
eliminate such terms, `t Hooft suggested using a modified gauge fixing
Lagrangian \cite{tHooft:1971rn,Fujikawa:1972fe}
\be\label{eq3}
{\cal L}^{(1)} = -\frac{1}{2\alpha}\left[\partial\cdot  V + 
\frac{ie\alpha}{2}\left(f^\star h -f h^\star \right)\right]^2 ;
\ee
in the sum of Eqs. \eqref{eq1} and \eqref{eq3} these cross terms
cancel provided $f$ is a constant. The same advantage arises if one
were to use the non-quadratic gauge fixing Lagrangian 
\be\label{eq4}
{\cal L}^{(2)} = -\frac{1}{2\alpha}\left[\partial\cdot
  V\right]\left[\partial\cdot V + {ie\alpha}\left(f^\star h -f h^\star \right)\right] 
\ee
without the introduction of terms quadratic in $h$, $h^\star$

Incorporation of the gauge fixing term of Eq. \eqref{eq3} in the
effective Lagrangian when using the path integral quantization
involves the Faddeev-Popov procedure
\cite{Faddeev:1967fc,Hooft:1971fh}, which must be extended in order to
accommodate the gauge fixing of Eq. \eqref{eq4}
\cite{Brandt:2007td}. We will briefly review this modification.  

The features of the path integral in which we are interested can be
illustrated by considering the integral over the components of an
$n$-dimensional vector $\vec h$ 
\be\label{eq5}
Z = \int {\rm d}\vec h \exp{(-\vec h^T\unm\vec h)}
=\frac{\pi^{\frac{n}{2}}}{(\det\unm)^{\frac{1}{2}}}
\ee
If $\unm$ is an $n\times n$ matrix such that
$\unm(\una \vec \theta)=0$ where $\vec\theta$ is
an arbitrary vector then the integral in Eq. \eqref{eq5} is
undefined. In order to excise the vanishing  eigenvalue of
$\unm$, we insert three factors of ``$1$" into \eqref{eq5} 
\begin{subequations}\label{eq6}
\bea\label{eq6a}
1=\int{\rm  d}\vec\theta_1\delta(\unf(\vec h+\alpha\una\vec\theta_1)-\vec p)\det(\alpha\unf\una)
\\ \label{eq6b}
1=\int{\rm  d}\vec\theta_2\delta(\ung(\vec h+\alpha\una\vec\theta_2)-\vec p)\det(\alpha\ung\una)
\eea
and
\bea\label{eq8}
1=\left(\alpha\pi\right)^{-n}\int{\rm d}\vec p\; {\rm d}\vec
q\exp{\left(-\frac{1}{\alpha}\vec p^T\unn\vec q\right)}\det{\unn}.
\eea
\end{subequations}
The shift
\be\label{eq9}
\vec h \rightarrow \vec h -\alpha \una \vec \theta_{1}
\ee
leads to (after integrating out the $\vec p$ and $\vec q$ variables)
\bea\label{eq10}
Z&=&\left(\frac{\alpha}{\pi}\right)^n
\int{\rm  d}\vec\theta_1\int{\rm  d}\vec\theta\int{\rm  d}\vec h
\det(\unf\una)
\det(\ung\una)
\nonumber \\
&\times& \det(\unn)
\exp\left[-\vec
    h^T\left(\unm+\frac{1}{\alpha}\unf^T\unn\ung\right)\vec h \right. 
\nonumber \\  & & 
\left. \qquad \qquad \qquad -\vec h^{T}\unf^T\unn\ung\una\vec\theta\right]
\eea 
where $\vec\theta=\vec\theta_{2} - \vec\theta_{1}$. One could shift
$\vec h$ to remove the ``cross term"  in $\vec h$ and $\vec\theta$ but
this shift would not be a ``gauge transformation" of the form of
\eqref{eq9}. 

The determinants in Eq. \eqref{eq10} can be exponentiated using the
standard Berezin integral 
\be\label{eq11}
\det\undertilde{B} = \int{\rm d}\vec c\;{\rm d}\vec{\bar{c}}
\;\exp(-\vec{\bar{c}}^{\,T}\undertilde{B}\vec c)
\ee 
where $\vec c$, $\vec{\bar{c}}$ are Grassmann vectors; the first two
determinants in Eq. \eqref{eq10} lead to ``Faddeev-Popov" like ghosts
and the third to a ``Nielsen-Kallosh'' ghosts \cite{Nielsen:1978mp,Kallosh:1978de}  
which we will subsequently ignore, taking $\unn=\mbox{const.}$. The field
$\vec\theta$ is a ``Bosonic'' ghost. The integral over
$\vec\theta_{1}$  in Eq. \eqref{eq10} parametrizes the divergence in
Eq. \eqref{eq5} and can be absorbed into the normalization of $Z$. 

In the model of Eq. \eqref{eq1}, we make the identification
\be\label{eq12}
\vec h = (V_{\mu}, h, h^\star)^{T}
\ee
so that
\begin{equation}\label{eq13}
\una=\left(\partial_{\mu},ie(f+h),-ie(f^\star+h^\star)\right)^{T}.
\end{equation}
From Eqs. \eqref{eq4}, \eqref{eq6} and \eqref{eq9} $\vec\theta$ is a scalar and
\begin{subequations}\label{eq14}
\bea
\label{eq14a}
\unf&=&\left(\partial_{\mu},0,0\right),
\\ \label{eq14b}
\ung&=&\left(\partial_{\mu}, i e \alpha f^\star, -i e \alpha f\right),
\\ \label{eq14c}
 \unn&=&\frac{1}{2}
\eea
\end{subequations}
so that the ghost actions are
\begin{equation}\label{eq15}
{\cal L}_{b}={\bar{b}} (\partial^2) b
\end{equation}
and
\begin{equation}\label{16}
{\cal L}_{c}=\bar{c} \left[\partial^2-2\alpha e^{2} f^\star f - \alpha
  e^2(f^\star h + f h^\star)\right]c 
\end{equation}
and the argument of the exponential in Eq. \eqref{eq10} is
\be\label{eq17}
{\cal L}_{T}={\cal L}_{Cl}+{\cal L}^{(2)}_{gf} 
-V^\mu \partial_{\mu} \left[\partial^2-\alpha e^2(2f^\star f 
+ f^\star h + f h^\star)\right]
\theta 
\ee

We now examine the global invariance possessed by the effective Lagrangian
\begin{equation}\label{}
{\cal L}_{eff}={\cal L}_{T}+{\cal L}_{b}+{\cal L}_{c} .
\end{equation}

\section{Global Gauge Invariance}
It is actually simpler to discuss the effective action before the
fields $\vec p$ and $\vec q$ in Eq. \eqref{eq8} are eliminated and
without making the shift of Eq. \eqref{eq9} so that 
\begin{eqnarray}\label{eq19}
&{\cal L}_{eff} = -\vec h^{T}\unm\vec h+\dfrac{1}{\alpha}\left[
\vec p^{\;T}\unf(\vec h+\alpha\una\vec\theta_1)\right.&
\nonumber\\ 
&\qquad \left. +\vec q^{\;T}\ung(\vec h+\alpha\una\vec\theta_2)
+\vec p^{\;T}\unn^{-1} \vec q\right]&
\nonumber\\ 
&+\vec{\bar{b}}\unf\una\vec b+\vec{\bar{c}}\ung\una\vec c
+\vec{\bar{k}}\unf\una\vec k&
\end{eqnarray}
Following the usual BRST transformation, we begin with 
\begin{subequations}\label{eq20}
\begin{eqnarray}\label{eq20a}
\delta \vec h &=& \una(\xi\vec c+\zeta \vec d)\epsilon
\\ \label{eq20b}
\delta\vec p&=&\delta\vec q=0=\delta\vec{\bar{k}}=\delta\vec k
\end{eqnarray}
\end{subequations}
where $\epsilon$ is a Grassmann constant and $\xi$  
and $\zeta$ are ordinary constants. 
If now the ghost fields are
also varied then the effective Lagrangian also undergoes the change 
\bea\label{eq21}
\delta{\cal L}_{eff}&=&\frac{1}{\alpha}\left[
\vec p^{\;T}\unf(\delta\vec
h+\alpha\una\delta\vec\theta_1+\alpha{\una}{}_{,l}\vec\theta_1\delta h_l)\right. 
\nonumber \\  & & \left.
\;\;\;+\vec q^{\;T}\ung(\delta\vec
h+\alpha\una\delta\vec\theta_2+\alpha{\una}{}_{,l}\vec\theta_2\delta h_l)\right]
\nonumber \\ & &\;\;\;+
\delta\vec{\bar{b}}\unf\una\vec b +\vec{\bar{b}}\unf\una{}_{,l}\delta h_l\vec b 
+\vec{\bar{b}}\unf\una\delta \vec b 
\nonumber \\ & &\;\;\;+
\delta\vec{\bar{c}}\ung\una\vec c +\vec{\bar{c}}\ung\una{}_{,l}\delta h_l\vec c 
+\vec{\bar{c}}\ung\una\delta \vec c 
\eea
We have used the fact that the classical Lagrangian is unaltered under
the transformation of Eqs. \eqref{eq20} and assumed that
$\unn$ is an invariant. In order to ensure that in
Eq. \eqref{eq21} $\delta{\cal L}_{eff}=0$, Eq. \eqref{eq20} must be supplemented by 
\begin{subequations}\label{eq22}
\bea\label{eq22a}
\delta\vec\theta_1&=&-\frac{\zeta}{\alpha}\vec c \;\epsilon,
\\ \label{eq22b}
\delta\vec\theta_2&=&-\frac{\xi}{\alpha}\vec b \;\epsilon,
\eea
\end{subequations}
\begin{subequations}\label{eq23}
\bea\label{eq23a}
\delta b_i&=&\frac{1}{2}\xi f_{ab;i} b_a b_b \;\epsilon,
\\ \label{eq23b}
\delta c_i&=&\frac{1}{2}\zeta f_{ab;i} c_a c_b \;\epsilon,
\eea
\end{subequations}
\begin{subequations}\label{eq24}
\bea
\label{eq24a}
\delta \bar{b}_i-\xi p_i\epsilon -\alpha\xi(\vec p^{\;T}\unf\una{}_{,l}\vec\theta_1)(\una\unl{}^{-1})_{li}\epsilon
\nonumber \\
-\alpha\xi(\vec q^{\;T}\ung\una{}_{,l}\vec\theta_2)(\una\unl{}^{-1})_{li}\epsilon
\nonumber \\
-\xi(\vec{\bar{c}}^{\;T}\ung\una{}_{,l}\vec c)(\una\unl{}^{-1})_{li}\epsilon=0,
\\
\label{eq24b}
\delta \bar{c}_i-\zeta q_i\epsilon -\alpha\zeta(\vec q^{\;T}\ung\una{}_{,l}\vec\theta_2)(\una\uns{}^{-1})_{li}\epsilon
\nonumber \\
-\alpha\zeta(\vec p^{\;T}\unf\una{}_{,l}\vec\theta_1)(\una\uns{}^{-1})_{li}\epsilon
\nonumber \\
-\zeta(\vec{\bar{b}}^{\;T}\unf\una{}_{,l}\vec b)(\una\uns{}^{-1})_{li}\epsilon=0,
 \\
(\unl\equiv\unf\una,\;\;\uns\equiv\ung\una)\nonumber.
\eea
\end{subequations}
In Eqs. \eqref{eq23a} and \eqref{eq23b} we have used the identity
\begin{equation}\label{eq28}
A_{ab,c}A_{cd} - A_{ad,c}A_{cb}=f_{bd;c}A_{ac}.
\end{equation}
This follows from the fact that the commutator of two gauge
transformations must be a gauge transformation. 

If the shift of Eq. \eqref{eq9} were to occur, then in
Eqs. \eqref{eq19}, \eqref{eq20}, \eqref{eq21}, \eqref{eq22},
\eqref{eq23} and \eqref{eq24} one sets $\vec\theta_{1}=\zeta=0$, $\vec\theta_{2}=\theta$.

Because of the presence of the operators $\undertilde{\Lambda}^{-1}$,
$\undertilde{\Sigma}^{-1}$ in Eqs. \eqref{eq24a} and \eqref{eq24b} these
transformations are non-local. With the identifications of
Eqs. \eqref{eq14}, the transformations of Eqs. 
\eqref{eq22}, \eqref{eq23} and \eqref{eq24} become 
\begin{widetext}
\begin{subequations}\label{eq29}
\bea\label{eq29a}
\delta \theta_1&=&-\frac{\zeta}{\alpha} c \,\epsilon,\qquad
\delta \theta_2=-\frac{\xi}{\alpha} b \,\epsilon,
\\ \label{eq29b}
\delta b&=&\delta c = 0,
\\ \label{eq29c}
\delta{\bar{b}}&=&\xi p \epsilon-i\alpha e^3 \partial^{-2}
\left[\alpha\xi q(f^\star h - f h^\star)\theta_2+\bar c
(f^\star h-f h^\star) c\right]\,\epsilon ,
\\ \label{eq29d}
\delta{\bar{c}}&=&\xi q \epsilon-i\alpha e^3 
\left[\partial^{2}-2\alpha e^2 f^\star f -\alpha(f^\star h + f h^\star)\right]^{-1}
\left[\alpha\zeta p(f^\star h - f h^\star)\theta_1+\bar b 
(f^\star h-f h^\star) b\right]\,\epsilon.
\eea
\end{subequations}
\end{widetext}
These transformations can be used to find relationships between
Green's functions in much the same way that the usual BRST
transformations can be employed. 

\section{Discussion}
We have demonstrated that non-quadratic gauge fixing terms can be used
in spontaneously broken theories and that these give rise to two
Fermionic and one Bosonic ghosts. The full effective action possess a
global gauge invariance with  a constant Grassmann gauge parameter. This
ensures that the transformation is a cohomology. 

The existence of a global gauge invariance in the effective action
introduced above may make it possible to devise a quantization
procedure similar to the ``BRST quantization''  procedure of Refs. 
\cite{Curci:1976yb,Kugo:1977yx}. This would guarantee that the theory 
is unitary and ghost-free, an interesting problem especially if one
wants to use the TT-gauge discussed in Ref. \cite{Brandt:2007td}.

\begin{acknowledgments}
F. T. Brandt would like to thank Fapesp and CNPq for financial support.
D. G. C. McKeon would like to thank R. MacLeod for a helpful suggestion.
\end{acknowledgments}


\end{document}